\documentclass[%
aps,nofootinbib,notitlepage,superscriptaddress,10Limberkipt,prd, twocolumn,
 amsmath,amssymb
]{revtex4-2}

\usepackage[caption=false]{subfig}
\usepackage{braket}
\usepackage{float}
\usepackage{lipsum}
\usepackage{graphicx}
\usepackage{dcolumn}
\usepackage{bm}
\usepackage{natbib}
\usepackage{hyperref}
\hypersetup{
	colorlinks = true,
    linkcolor = Maroon,
    urlcolor  = Maroon,
    citecolor = Maroon
}
\usepackage{afterpage}
\usepackage{placeins}
\usepackage{microtype}
\usepackage{verbatim}
\usepackage[amssymb]{SIunits}
\usepackage{tabularx}
\usepackage[dvipsnames]{xcolor}
\usepackage{tikz}

\usepackage{color}
\definecolor{darkgreen}{RGB}{0,120,0}

\newcommand{\vk}{{\bm k}}
\newcommand{\vq}{{\bm q}}

\newcommand{\av}[1]{\langle{#1}\rangle}

\newcommand{\threej}[6]{ \begin{pmatrix}
  #1 & #2 & #3 \\
  #4 & #5 & #6 
 \end{pmatrix}}

\newcommand{\hatbn}{{\hat{\boldsymbol n}}}

\def\gsim{ \lower .75ex \hbox{$\sim$} \llap{\raise .27ex \hbox{$>$}} }
\def\lsim{ \lower .75ex \hbox{$\sim$} \llap{\raise .27ex \hbox{$<$}} }

\def\dalam{\hbox
{\vrule\vbox{\hrule\hbox to 1ex{ \hfill}\kern 1 ex\hrule}\vrule}}

\def\1/2{\hbox{$ {1 \over 2}$ }}

\def\h{\hbar}
\def\i/h{{i \over \h}}

\begin{document}

\title{Consistently constraining \texorpdfstring{$\bm f_{\rm NL}$}{} with the squeezed lensing bispectrum using consistency relations}

\author{Samuel~Goldstein}
\email{sjg2215@columbia.edu}
\affiliation{Center for Theoretical Physics, Department of Physics, Columbia University, New York, NY 10027, USA}

\author{Oliver~H.\,E.~Philcox}
\affiliation{Simons Society of Fellows, Simons Foundation, New York, NY 10010, USA}
\affiliation{Center for Theoretical Physics, Department of Physics, Columbia University, New York, NY 10027, USA}

\author{J.\,Colin~Hill}
\affiliation{Center for Theoretical Physics, Department of Physics, Columbia University, New York, NY 10027, USA}

\author{Angelo~Esposito}
\affiliation{Dipartimento di Fisica, Sapienza Universit\`a di Roma, Piazzale Aldo Moro 2, I-00185 Rome, Italy}
\affiliation{INFN Sezione di Roma, Piazzale Aldo Moro 2, I-00185 Rome, Italy}

\author{Lam~Hui}
\affiliation{Center for Theoretical Physics, Department of Physics, Columbia University, New York, NY 10027, USA}

\begin{abstract}
We introduce a non-perturbative method to constrain the amplitude of local-type primordial non-Gaussianity ($f_{\rm NL}$) using squeezed configurations of the CMB lensing convergence and cosmic shear bispectra. First, we use cosmological consistency relations to derive a model for the squeezed limit of angular auto- and cross-bispectra of lensing convergence fields in the presence of $f_{\rm NL}$. Using this model, we perform a Fisher forecast with specifications expected for upcoming CMB lensing measurements from the Simons Observatory and CMB-S4, as well as cosmic shear measurements from a Rubin LSST/{\it Euclid}-like experiment. Assuming a minimum multipole $\ell_{\rm min}=10$ and maximum multipole $\ell_{\rm max}=1400$, we forecast $\sigma_{f_{\rm NL}}=175$ ($95$) for Simons Observatory (CMB-S4). Our forecasts improve considerably for an LSST/{\it Euclid}-like cosmic shear experiment with three tomographic bins and $\ell_{\rm min}=10$ and $\ell_{\rm max}=1400$ ($5000$) with $\sigma_{f_{\rm NL}}=31$ ($16$). A joint analysis of CMB-S4 lensing and LSST/{\it Euclid}-like shear yields little gain over the shear-only forecasts; however, we show that a joint analysis could be useful if the CMB lensing convergence can be reliably reconstructed at larger angular scales than the shear field. The method presented in this work is a novel and robust technique to constrain local primordial non-Gaussianity from upcoming large-scale structure surveys that is completely independent of the galaxy field (and therefore any nuisance parameters such as $b_\phi$), thus complementing existing techniques to constrain $f_{\rm NL}$ using the scale-dependent halo bias. 
\end{abstract}

\maketitle

\section{Introduction}\label{Sec:Intro}

One of the primary goals of ongoing and upcoming galaxy surveys such as the Dark Energy Spectroscopic Instrument (DESI) \cite{DESI:2016fyo}, {\it Euclid} \cite{Amendola:2016saw}, {\it SPHEREx} \cite{Dore:2014cca}, and the Vera C.~Rubin Observatory Legacy Survey of Space and Time (LSST)~\cite{LSSTScience:2009jmu} is to reveal information about the physics behind the primordial perturbations that evolved into present-day cosmic structures. In the standard cosmological model, these perturbations are produced during an inflationary epoch in which the Universe underwent a period of rapid accelerated expansion. The simplest single-field models of inflation predict initial conditions that are almost perfectly Gaussian and adiabatic~\cite{Maldacena:2002vr,Creminelli:2004yq,Creminelli:2011rh, Pajer:2013ana}; however, a wealth of more complex models exist that predict departures from Gaussianity~\cite{Meerburg:2019qqi,Achucarro:2022qrl}. As such, searches for primordial non-Gaussianity (PNG) can powerfully probe the physics of the early Universe.

Currently, the tightest constraints on a wide range of parameters characterizing the amplitude of PNG in various shapes come from analysis of the cosmic microwave background (CMB)~\cite{Planck:2019kim}; nevertheless, large-scale structure (LSS) observations provide a complementary approach to competitively constrain PNG~\cite{DAmico:2022gki, Cabass:2022wjy, Cabass:2022ymb, McCarthy:2022agq, Cabass:2022epm, DESI:2023duv, Rezaie:2023lvi}. For local-type PNG, the subject of this work, LSS constraints are typically derived by taking advantage of its distinct imprint on halo clustering via the ``scale-dependent bias". This effect manifests as an enhancement in the amplitude of the large-scale power spectrum of biased tracers relative to the expectation from Gaussian initial conditions~\cite{Dalal:2007cu, Matarrese:2008nc, Slosar:2008hx, Desjacques:2008vf}. Constraining the amplitude of local PNG, $f_{\rm NL}$,\footnote{Throughout this paper, we use $f_{\rm NL}\equiv f_{\rm NL}^{\rm loc}$.} using the scale-dependent bias is a key goal of upcoming surveys. The potential of this technique was recently demonstrated in an analysis of DESI photometric clustering data, which provided the most precise LSS constraint on $f_{\rm NL}$ to date~\cite{Rezaie:2023lvi}. However, the impact of foreground and systematic effects on this result remains unclear.

A potential limitation of using the scale-dependent bias to constrain local PNG is that the derived constraints on $f_{\rm NL}$ require precise knowledge of the impact of local PNG on galaxy formation, due to the perfect degeneracy between $f_{\rm NL}$ and the non-Gaussian halo bias parameter $b_\phi$~\cite{Reid_2010, Barreira:2020kvh, Barreira:2021ueb, Lazeyras:2022koc,Barreira:2022sey, Barreira:2023rxn}. This situation motivates the development of alternative methods to constrain local PNG using LSS that do not rely on the scale-dependent bias. A promising option is to instead use the weak lensing bispectrum, since this is sensitive to the (unbiased) total matter distribution.

The prospect of constraining local PNG using the weak lensing bispectrum was first discussed in~\citet{Takada:2003ef} (see also~\citep{2012MNRAS.421..797S, PhysRevD.83.123005, Grassi:2013ana}), where the authors found that weak lensing convergence bispectrum tomography does not yield competitive constraints on $f_{\rm NL}$. The goal of our paper is to revisit the feasibility of this approach and investigate whether the situation has improved given the significant advancements in both modeling and observations over the past two decades. Our work builds upon Ref.~\cite{Takada:2003ef} in several ways. Firstly, while Ref.~\cite{Takada:2003ef} used a tree-level perturbation theory model for the local PNG contribution to the lensing bispectrum, we utilize a recently developed non-perturbative model for the matter bispectrum based on the LSS consistency relations~\cite{Peloso:2013zw, Kehagias:2012pd} that has been validated deep into the non-linear regime using $N$-body simulations~\cite{Esposito:2019jkb, Goldstein:2022hgr}. This model enables us to include non-linear modes in our analysis, leveraging the unprecedented depth of ongoing and upcoming weak lensing experiments. Secondly, we use realistic galaxy source distributions and number densities expected for an LSST/{\it Euclid}-like survey to directly forecast the constraining power of Stage-IV shear experiments. Finally, we forecast the constraining power of CMB lensing bispectrum measurements using lensing reconstruction noise properties expected from the imminent Simons Observatory~\cite{SimonsObservatory:2018koc} and future CMB-S4~\cite{CMB-S4:2016ple} experiments. To our knowledge, this is the first forecast for constraining PNG using the CMB lensing convergence bispectrum, which has previously been shown to be a promising probe of cosmology~\cite{Namikawa:2016jff}.

The remainder of the paper is organized as follows. In Section~\ref{Sec:theory_background}, we provide theoretical background on local PNG and weak lensing and present expressions for the convergence power spectra and bispectra, as well as their covariances. In Section~\ref{Sec:forecast_setup}, we discuss the forecast setup, before presenting the corresponding results in Section~\ref{Sec:results}. Section~\ref{Sec:conclusions} summarizes our conclusions and the Appendix contains a discussion of the impact of non-linear effects on our forecasts.

\vskip 4pt
\noindent{\it Conventions:} Throughout this paper we work in natural units, $c=1$. We assume a fiducial spatially flat $\Lambda$CDM cosmology based on the \emph{Planck} 2018 results~\cite{Planck:2018vyg} with $\Omega_{m} = 0.311$, $\Omega_b = 0.0493$,  $h = 0.674$, $n_s = 0.965$, $\sigma_8 = 0.811$, and $\tau=0.054$.  We assume three species of massless neutrinos.

\section{Theoretical background}\label{Sec:theory_background}

To forecast how well the squeezed lensing bispectrum can constrain $f_{\rm NL}$, we require a model for the lensing convergence bispectrum (and its covariance) in the presence of local PNG. We derive these results in this section. We first derive a non-perturbative expression for the unequal-time squeezed 3D matter bispectrum in the presence of $f_{\rm NL}$ based on Ref.~\cite{Goldstein:2022hgr}. We then introduce the weak lensing convergence field and compute expressions for the convergence power spectrum and bispectrum and their covariances in terms of the 3D matter power spectrum and bispectrum. Some of the results of this section are standard in the lensing literature~\cite{ Bartelmann:1999yn, Lewis:2006fu, Bartelmann:2016dvf, Kilbinger:2014cea}; nevertheless, we include them here both for the sake of completeness and to highlight the assumptions underlying our forecasts.

\subsection{Conventions}

We first establish some notation. For an overdensity field $\delta(\bm q)$, the 3D power spectrum is defined by
\begin{equation}
    \langle \delta(\bm k)\delta(\bm k') \rangle \equiv (2\pi)^3\delta^D(\bm k+\bm k')\,P(k, z_k) \,, 
\end{equation}
where we have explicitly included the time-dependence of $\delta(\bm k)$. Similarly, the 3D bispectrum is defined by
\begin{equation}
    \langle \delta(\bm q) \delta(\bm k) \delta(\bm p) \rangle \equiv (2\pi)^3\delta^D(\bm q+\bm k+\bm p)\,B(\bm q, \bm k, z_q, z_k, z_p) \,.
\end{equation}
Here, we are interested in the squeezed limit of the bispectrum, where one of the modes is much smaller than the other two, {i.e.}, $q\ll k\simeq p$. In this limit, the contributions to the lensing bispectrum satisfy $z_k\simeq z_p$; therefore, we use the notation $B(\bm q, \bm k, z_q, z_k)$ to indicate the unequal-time bispectrum (fixing $z_p=z_k$). Finally, when working with correlators of the lensing convergence, we will often specify the time-dependence implicitly in terms of the comoving distance, $\chi$, to redshift $z$.

\subsection{Squeezed matter bispectrum in the presence of local primordial non-Gaussianity}

To derive an expression for squeezed configurations of the lensing bispectrum, we first require a model for the unequal-time squeezed matter bispectrum in the presence of local PNG. Our derivation follows that of~\cite{Goldstein:2022hgr} (see also \citep{Giri:2022nzt,Giri:2023mpg}), but is generalized to include unequal-time correlations between the long and short modes. 

Local PNG is parametrized by a primordial gravitational potential on sub-horizon scales given by~\citep[e.g.,][]{Komatsu:2001rj,Scoccimarro:2011pz}
\begin{align}\label{eq:fNL_def}
    \begin{split}
       \Phi(\bm x) = \Phi_G(\bm x) + f_{\rm NL}\big(\Phi_G^2(\bm x) - \langle \Phi_G^2\rangle  \big) \,,
    \end{split}
\end{align}
where $\Phi_G(\bm x)$ is a Gaussian random field. To determine the squeezed bispectrum we evaluate the correlator $\av{\delta(\bm q) P(k,\chi_k|\Phi_L)}$ where $\delta(\bm q)$ is the soft mode density field and $P(k,\chi_k|\Phi_L)$ is the locally measured small-scale power spectrum in the presence of a background long-wavelength potential $\Phi_L$~\cite{Giri:2022nzt}. This can be expanded as
\begin{align}
    \begin{split}
        P(k,\chi_k|\Phi_L) ={} &P(k,\chi_k|0)\\
        &+\int d{\vq^\prime} \, \frac{\partial P(k,\chi_k)}{\partial\,\Phi_{L}(\vq^\prime)}\, \Phi_{L}(\vq^\prime) + \dots \,,
    \end{split}
\end{align}
where we are assuming that $q^\prime\equiv |\vq'| \ll k$, such that we can treat the long mode as a background, in the presence of which the power spectrum is evaluated. This induces a coupling between the hard mode power spectrum, $P(k,\chi_k)$, and the soft mode density field, $\delta(\vq)$.  The resulting squeezed limit bispectrum is given by
\begin{align}\label{eq: bspec_def}
    \begin{split}
        B(\vq,\vk,\chi_q,\chi_k) &= \int d{\vq^\prime} \frac{\partial P(k,\chi_k)}{\partial\,\Phi_{L}(\vq^\prime)}\av{\delta(\vq) \Phi_{L}(\vq^\prime)} \\
        &=\frac{\partial P(k,\chi_k)}{\partial\,\Phi_{L}(\vq)}\frac{P(q,\chi_q)}{\alpha(q,\chi_q)} \,.
    \end{split}
\end{align}
Here, we have used Poisson's equation to relate the long-wavelength density and potential fields, $\delta(\bm q) =\alpha(q,\chi_q)\,\Phi_L(\vq)$, with
\begin{align} \label{eq: alpha-q-def}
	\begin{split}
		\alpha(q,\chi_q) \equiv \frac{2}{3}\frac{q^2T(q)D_{\rm md}(\chi_q)}{\Omega_{m}H_0^2} \,,
	\end{split}
\end{align}
where $\Omega_m$ and $H_0$ are the matter density and expansion rate today, $T(q)$ is the transfer function (normalized to unity for $q \to 0$), and $D_{\rm md}(z)$ is the growth factor, normalized to $a(z)=1/(1+z)$ in the matter-dominated era.

We evaluate the potential derivative using the separate Universe formalism leading to the following expression for the primordial contribution to the late-time matter bispectrum in the squeezed limit~\cite{Goldstein:2022hgr, Giri:2023mpg}:
\begin{align} \label{eq: sq_Bk_fNL_3Dmatter}
	\begin{split}
		B_{\rm prim}(\vq,\vk,\chi_{q},\chi_{k}) ={}& \frac{6f_{\rm NL}\Omega_mH_0^2}{D_{\rm md}(\chi_q)} \, \frac{\partial P(k,\chi_k)}{\partial\log\sigma_8^2} \, \\ 
  &\times \frac{P(q,\chi_q)}{q^2 T(q)}+\mathcal{O}\big(f_{\rm NL}^2\big) \,.
	\end{split}
\end{align} 
 Notice that the soft-mode dependence of the primordial contribution to the squeezed matter bispectrum scales as $P(q,\chi_q)/q^2$; however, the LSS consistency relations~\cite{Peloso:2013zw, Kehagias:2012pd} ensure that, in the absence of local PNG and equivalence-principle-violating physics, the ratio $B(\bm q, \bm k, \chi_q,\chi_k)/P(q,\chi_q)$ has no $1/q$ and $1/q^2$ poles in the squeezed limit $q/k\rightarrow 0$
 (see also \cite{Esposito:2019jkb}).\footnote{In the unequal-time limit in the hard modes (i.e., $\chi_{k_1}\neq \chi_{k_2}$), the squeezed bispectrum has a term proportional to $(D_{\rm md}(\chi_{k_1})-D_{\rm md}(\chi_{k_2}))P(q,\chi_q)/q$; however, this contribution vanishes in the limit $\chi_{k_2}\to \chi_{k_3}$, which is assumed here.} 
 It is worth stressing that this statement about the lack of poles, in the absence of local PNG and equivalence-principle-violation, is robust: it holds even if the high momentum ($k$) modes are in the nonlinear regime, and even if they are affected by baryonic feedback processes \cite{Horn:2014rta,Horn:2015dra}.
 As such, we can split the late-time squeezed matter bispectrum into a primordial contribution and a gravitational contribution $B=B_{\rm prim}+B_{\rm grav}$, where the primordial contribution is, up to $\mathcal{O}(f_{\rm NL}^2),$ described by Eq.~\eqref{eq: sq_Bk_fNL_3Dmatter}.

 The gravitational term $B_{\rm grav}$ is difficult to model beyond perturbative scales; nevertheless, as shown in~\cite{Esposito:2019jkb}, the squeezed bispectrum is well-described by a power series 
\begin{align} 
	\begin{split}
    B_{\rm grav}(\vq,\vk, \chi_q,\chi_k)={}& \sum\limits_{n=0}^\infty a_n(k, \theta,\chi_k)\bigg(\frac{q}{k}\bigg)^n \\
    & \times P(q,\chi_q)P(k, \chi_k) \,,
    \end{split}
\end{align} 
where the $a_i$'s are coefficients characterizing the response of small-scale matter clustering to a long-wavelength mode and $\theta$ is the angle between $\vq$ and $\vk$. As described in Refs.~\cite{Esposito:2019jkb, Goldstein:2022hgr}, by angular averaging over all available short modes that satisfy the triangle inequality, the odd-order coefficients vanish, and the remaining coefficients become independent of the angle $\theta$.  In practice, one truncates the series at some finite $n$. By comparing to $N$-body simulations, Ref.~\cite{Goldstein:2022hgr} found that truncating at $n=2$ is sufficient for a wide range of soft modes. The gravitational contribution to the matter bispectrum is then
\begin{align}\label{eq:b_3D_grav}
     B_{\rm grav}(\bm q,\bm k, z_q,z_k)=&\,a_0(k, \chi_k)P(k)P(q)+\\
     &\,{a}_2(k,\chi_k)\frac{q^2}{k^2}P(k)P(q) \,, \notag
\end{align}
which carries two scale- and redshift-dependent ``nuisance'' parameters, ${a}_0(k,\chi_k)$ and ${a}_2(k,\chi_k)$. In this work, we will ignore the ${a}_2(k,\chi_k)$ contribution because it is subdominant and largely uncorrelated with $f_{\rm NL}$~\cite{Goldstein:2022hgr}.\footnote{As shown in Appendix B of~\cite{Goldstein:2022hgr}, ${a}_2(k,\chi_k)=0$ is a valid assumption for a wide range of scale cuts.} Furthermore, for our fiducial forecasts, we will assume perfect knowledge of ${a}_0(k,\chi_k)$ via the angular averaged bispectrum consistency condition from~\cite{Valageas:2013zda, Nishimichi:2014jna},
\begin{align}\label{eq:a0_CR}
    {a}_0(k,\chi_k)=1+\frac{13}{21}\frac{\partial\log P(k,\chi_k)}{\partial \log D_{\rm md}(\chi_k)}-\frac{1}{3}\frac{\partial\log P(k,\chi_k)}{\partial \log k} \,.
\end{align}
Although Eq.~\eqref{eq:a0_CR} is non-perturbative and can therefore be applied to non-linear scales, it is expected to break down at small scales and low redshifts due to baryonic effects and departures from an Einstein--de Sitter universe~\cite{Valageas:2013zda, Nishimichi:2014jna}. Furthermore, even if Eq.~\eqref{eq:a0_CR} is valid over the scales and redshifts considered in this work, modeling the derivatives in Eq.~\eqref{eq:a0_CR} can be challenging. Consequently, we also consider forecasts for a more pessimistic scenario in which ${a}_0(k,\chi_k)$ is a free amplitude that we marginalize over, as was done in Ref.~\cite{Goldstein:2022hgr}.

\subsection{Weak gravitational lensing}\label{Subsec:weak_lens }

In this section, we introduce the weak lensing convergence field and derive theoretical predictions for the lensing convergence power spectrum and bispectrum and their associated covariances. For a detailed treatment of weak lensing see, e.g.,~\cite{ Bartelmann:1999yn, Lewis:2006fu, Bartelmann:2016dvf, Kilbinger:2014cea}. 

Assuming the Born approximation and working at linear order in the matter density fluctuation, the convergence field $\kappa^{(i)}(\hatbn)$ is a weighted projection of the matter density field:
\begin{equation}\label{eq:converge_generic}
    \kappa^{(i)}(\hatbn)=\int_0^{\chi_s}d\chi\,W^{(i)}(\chi)\,\delta_m(\chi\hatbn,\chi)\,,
\end{equation}
where $\hat{\bm n}$ is a unit vector, $\chi_s$ is the comoving distance to the photon source, and $W^{(i)}(\chi)$ is the projection kernel. The exact form of $W^{(i)}(\chi)$ is determined by the specifics of the lensing source.  For CMB lensing, the kernel is 
\begin{equation}\label{eq:CMB_lensing_window}
W^{\kappa_{\rm CMB}}(\chi)\equiv \frac{3\,H_0^2\,\Omega_m\chi}{2\,a(\chi)}\left( \frac{\chi_*-\chi}{\chi_*} \right),
\end{equation}
where $\chi_*$ is the comoving distance to the last-scattering surface at $z_*\simeq 1090$, and $a(\chi)$ is the scale factor. For cosmic shear, the kernel is
\begin{equation}
    W^{\kappa_{\rm g}, (i)}(\chi)=\frac{3\,H_0^2\,\Omega_m\chi}{2\,a(\chi)}\int_\chi^\infty\!\! d\chi^\prime \,p^{(i)}_s(\chi^\prime)\bigg(\frac{\chi^\prime-\chi}{\chi^\prime}\bigg) \,,
\end{equation}
where $p^{(i)}_s(\chi)$ is the redshift distribution of source galaxies in the $i$-th tomographic bin, satisfying the normalization condition $\int d\chi^\prime\, p_s^{(i)}(\chi^\prime)=1$.

It is convenient to expand the convergence field in spherical harmonics,
\begin{equation}\label{eq:conv_harm_expand}
    \kappa^{(i)}(\hatbn)=\sum\limits_{\ell,m}\kappa^{(i)}_{\ell m}Y_{\ell m}(\hatbn) \,.
\end{equation}
In the following sections, we will derive expressions for the auto- and cross-power spectra and bispectra of arbitrary convergence fields $\kappa^{(i)}_{\ell m}$ in terms of the 3D matter power spectrum and bispectrum.

In practice, galaxy weak lensing surveys measure cosmic shear instead of convergence; however, the convergence field can be reconstructed from shear measurements~\cite{Kaiser:1992ps,Castro:2005bg, Leistedt:2016dda, Wallis:2017lwt, DES:2017stf, DES:2019ujq, Barthelemy:2023mer}. The reconstructed convergence field is a (somewhat) biased estimate of the underlying convergence field; therefore, in a real analysis, one would likely directly model the shear bispectrum instead of the convergence bispectrum.  For CMB lensing, the convergence field itself can be directly reconstructed from the observed CMB temperature and polarization anisotropies~(e.g.,~\cite{Hu:2001kj}).

\subsubsection{Convergence power spectrum}

The angular power spectrum between two convergence fields, $C_\ell^{(ij)}$, is defined by $\big\langle \kappa_{\ell m}^{(i)}\kappa_{\ell',m'}^{*(j)}\big\rangle=\delta^K_{\ell\ell'}\delta^K_{mm'}C_\ell^{(ij)}$, assuming statistical isotropy. Using Eqs.~\eqref{eq:converge_generic} and~\eqref{eq:conv_harm_expand}, we can express $C_\ell^{(ij)}$ as
\begin{align} \label{eq:angular_power_spec_full}
	\begin{split}
   C_\ell^{(ij)} ={}& \frac{2}{\pi}\int\,d\chi_1\,W^{(i)}(\chi_1)\int d\chi_2\,W^{(j)}(\chi_2)\\
   &\times\int dk\,k^2 P(k,\chi_1, \chi_2)\,j_{\ell}(k\chi_1)\,j_{\ell}(k\chi_2) \,.
     \end{split}
 \end{align}
At high $\ell$, we can use the Limber approximation~\cite{Limber, Kaiser:1991qi, LoVerde:2008re} to replace the highly oscillatory spherical Bessel function by a Dirac delta function,
\begin{equation}\label{eq:Limber_approx}
    j_\ell(k\chi)\simeq \sqrt{\frac{\pi}{2\ell +1}}\delta^D\big(\ell + \tfrac{1}{2} -k\chi\big) \,, 
\end{equation}
leading to
\begin{equation}\label{eq:Cl_limber}
     C_{\ell\gg 1}^{(ij)}\simeq \int \frac{d\chi}{\chi^2}W^{(i)}(\chi)W^{(j)}(\chi)P\left(k=\frac{\ell+\frac{1}{2}}{\chi},\chi\right).
\end{equation}
In this work, we compute the angular power spectrum using the exact expression via the FFTLog~\cite{Hamilton:1999uv, Assassi:2017lea, Fang:2019xat} algorithm for $\ell<50$ and employ the Limber approximation for $\ell\geq 50$.

Assuming a fractional sky coverage $f_{\rm sky}$, the covariance between $C_\ell^{(ij)}$ and $C_{\ell'}^{(mn)}$ is~\cite{Takada:2003ef}
\begin{align} \label{eq:angular_power_spec_cov}
    {\rm{Cov}}\left[{C}^{(ij)}_\ell,{C}^{(mn)}_{\ell'}\right]={}&\frac{\delta^{\rm K}_{\ell\ell'}}{f_{\rm sky}(2\ell+1)}\\
    &\times\bigg({\tilde{C}}^{(im)}_\ell{\tilde{C}}^{(jn)}_\ell+{\tilde{C}}^{(in)}_{\ell}{\tilde{C}}^{(jm)}_\ell \bigg) \,, \notag
\end{align}
where we have neglected the connected non-Gaussian contribution and the super-sample covariance~\cite{Takada:2008fn, Sato:2009, Sato:2013mq, Takada:2013jwa, Krause:2016jvl, Barreira:2018jgd}. The connected non-Gaussian contribution is expected to be subdominant in Stage-IV convergence power spectra due to the suppression of non-Gaussianities in lensing, which projects quantities along the line-of-sight \cite{Takada:2008fn, Barreira:2018jgd}. Conversely, as shown in~\cite{Barreira:2018jgd}, the super-sample covariance can have a significant impact on the  convergence power spectra for Stage-IV shear surveys. Nevertheless, we ignore the non-Gaussian covariance so that we can determine the most optimistic forecasts for a given survey (and thus obtain an upper bound on the utility of our method). Note that in Eq.~\eqref{eq:angular_power_spec_cov}, we write ${\tilde{C}}^{(ij)}_\ell$ to emphasize that this is the \emph{observed} angular power spectrum, including the noise contribution as discussed in Sec.~\ref{Subsec:survey_specs}.

\subsubsection{Convergence bispectrum}\label{Sec:convergence_bispec}

The angular bispectrum $B_{\ell_1\ell_2\ell_3}^{(ijk)}$ is defined by~\citep[e.g.,][]{Hu:2000ee}
\begin{equation}
   \langle \kappa^{(i)}_{\ell_1m_1} \kappa^{(j)}_{\ell_2m_2} \kappa^{(k)}_{\ell_3m_3}\rangle \equiv \!{\footnotesize{\threej{\ell_1}{\ell_2}{\ell_3}{m_1}{m_2}{m_3}}} B^{(ijk)}_{\ell_1\ell_2\ell_3}  \,,
\end{equation}
where $\footnotesize{\threej{\ell_1}{\ell_2}{\ell_3}{m_1}{m_2}{m_3}}$ is the Wigner-3$j$ symbol. Using Eqs.~\eqref{eq:converge_generic} and~\eqref{eq:conv_harm_expand}, the convergence three-point function can be expressed as 
\begin{widetext}
    \begin{align}
        \begin{split}
	            \big\langle \kappa^{(1)}_{\ell_1m_1} \kappa^{(2)}_{\ell_2m_2} \kappa^{(3)}_{\ell_3m_3}\big\rangle=\frac{8}{\pi^3}\mathcal{G}^{\ell_1\ell_2\ell_3}_{m_1m_2m_3}\int dr\, r^2\bigg[\prod_{i=1}^3\, d\chi_i\,dk_i\,k_i^2\,W^{(i)}(\chi_i)j_{\ell_i}(k_i\chi_i)\,j_{\ell_i}(k_ir) \bigg]B(k_1,k_2,k_3,\chi_1,\chi_2,\chi_3) \,,
        \end{split}
    \end{align}
\end{widetext}
where we have introduced the Gaunt factor, $\mathcal{G}^{\ell_1\ell_2\ell_3}_{m_1m_2m_3} \equiv \footnotesize{\sqrt{\frac{(2\ell_1+1)(2\ell_2+1)(2\ell_3+1)}{4\pi}}\threej{\ell_1}{\ell_2}{\ell_3}{0}{0}{0}\threej{\ell_1}{\ell_2}{\ell_3}{m_1}{m_2}{m_3}}$. Again asserting isotropy, it is convenient to define the reduced bispectrum,  $\big\langle \kappa^{(i)}_{\ell_1m_1} \kappa^{(j)}_{\ell_2m_2} \kappa^{(k)}_{\ell_3m_3}\big\rangle\equiv\mathcal{G}^{\ell_1\ell_2\ell_3}_{m_1m_2m_3}b^{(ijk)}_{\ell_1\ell_2\ell_3}\,$, such that the reduced bispectrum and angular bispectrum are related by 
\begin{align}
    \begin{split}
        B_{\ell_1\ell_2\ell_3}^{(ijk)}&=\footnotesize{\sqrt{\tfrac{(2\ell_1+1)(2\ell_2+1)(2\ell_3+1)}{4\pi}}}\threej{\ell_1}{\ell_2}{\ell_3}{0}{0}{0}b_{\ell_1\ell_2\ell_3}^{(ijk)} \\
        &\equiv h_{\ell_1\ell_2\ell_3}\,b^{(ijk)}_{\ell_1\ell_2\ell_3}\,.
    \end{split}
\end{align}

As in Eq.~\eqref{eq:Cl_limber}, in the squeezed limit, which corresponds to $\ell_1\ll \ell_2\simeq \ell_3$, we can use the Limber approximation for the spherical Bessel functions $j_{\ell_2}(k_2\chi_2)$ and  $j_{\ell_3}(k_3\chi_3)$. This allows the reduced bispectrum to be written as
\begin{align} \label{eq:reduced_bispec_squeezed}
    b^{(ijk)}_{\ell_1\ell_2\ell_3}={}&\frac{2}{\pi}\int d\chi_1\,d\chi_2\left(\frac{W^{(i)}(\chi_1)W^{(j)}(\chi_2)W^{(k)}(\chi_2)}{\chi_2^2}\right) \notag \\
    &\times\int dq\,q^2 j_{\ell_1}(q\chi_1)j_{\ell_1}(q\chi_2)\\
    &\times B\left(q, \frac{\ell_2+\frac{1}{2}}{\chi_2}, \frac{\ell_3+\frac{1}{2}}{\chi_2}, \chi_1,\chi_2,\chi_2\right) \,. \notag
\end{align}
Notice that the assumption that $\chi_2\simeq\chi_3$ is exact in the Limber approximation.
If we also assume the Limber approximation for $\ell_1$, then 
\begin{align}
\begin{split}\label{eq:reduced_bispec_limber}
 b^{(ijk)}_{\ell_1\ell_2\ell_3}\simeq{}& \int d\chi\frac{W^{(i)}(\chi)W^{(j)}(\chi)W^{(k)}(\chi)}{\chi^4}\\ 
&\times B\left(\frac{\ell_1+\frac{1}{2}}{\chi}, \frac{\ell_2+\frac{1}{2}}{\chi}, \frac{\ell_3+\frac{1}{2}}{\chi},\chi\right)\,.
\end{split}
\end{align}
Here, we compute the squeezed angular bispectrum using Eq.~\eqref{eq:reduced_bispec_squeezed} via the FFTLog algorithm for $\ell_1<50$ and use the Limber approximation in Eq.~\eqref{eq:reduced_bispec_limber} for $\ell_1\geq 50.$

Finally, assuming $\ell_1\leq\ell_2\leq\ell_3$, the Gaussian covariance between the reduced bispectra $b_{\ell_1\ell_2\ell_3}^{(abc)}$ and $b_{\ell_1'\ell_2'\ell_3'}^{(ijk)}$ is
\begin{align}\label{eq:reduced_bspec_cov}
{\rm Cov}\left[b_{\ell_1\ell_2\ell_3}^{(abc)},
b_{\ell_1^\prime \ell_2^\prime \ell_3^\prime}^{(ijk)}\right]
&\simeq ~ h_{\ell_1\ell_2\ell_3}^{-1}h_{\ell_1'\ell_2'\ell_3'}^{-1}\frac{\Delta(\ell_1,\ell_2,\ell_3)}{f_{\rm sky}}\\
&\,\times\,
\tilde{C}^{(ai)}_{\ell_1}\tilde{C}^{(bj)}_{\ell_2}\tilde{C}^{(ck)}_{\ell_3}\delta^K_{\ell_1\ell_1^{\prime}}\delta^K_{\ell_2\ell_2^{\prime}}\delta^K_{\ell_3\ell_3^{\prime}} \,, \nonumber
\end{align}
where $\Delta(\ell_1,\ell_2,\ell_3)$ is a symmetry factor that is equal to six for equilateral triangles, two for isosceles triangles, and one otherwise. As in Eq.~\eqref{eq:angular_power_spec_cov}, Eq.~\eqref{eq:reduced_bspec_cov} includes only the Gaussian contribution to the covariance. Although non-Gaussian terms can have a significant impact on the lensing convergence bispectrum covariance~\cite{Kayo_2012, Kayo:2013aha, Sato:2013mq, Chan:2017fiv, Floss:2022wkq, Bayer:2022nws}, especially for squeezed configurations, we neglect them in order to forecast the most optimistic possible constraints on $f_{\rm NL}$ achievable with the presented method.

\section{Forecast setup}\label{Sec:forecast_setup}

\begin{figure*}[!tbp]
\includegraphics[width=\linewidth]{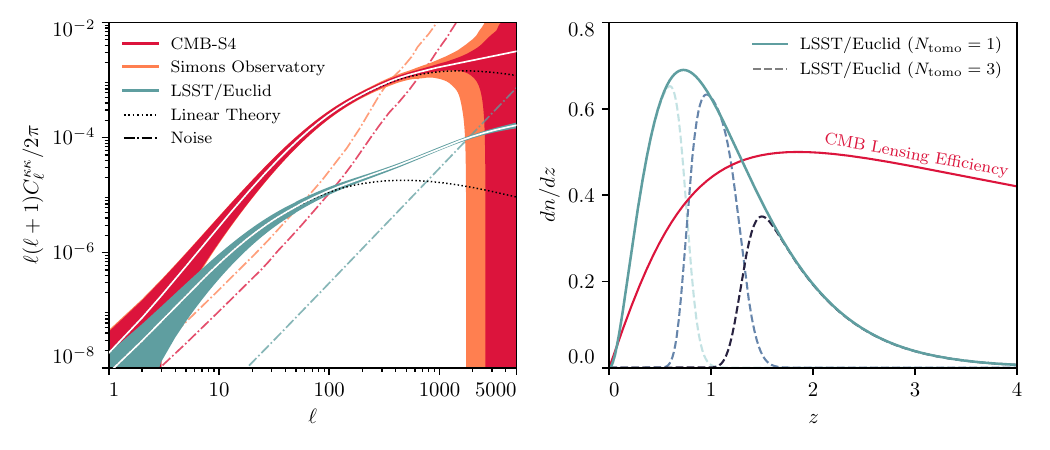}
\caption{
\textbf{Left}: Angular auto-power spectra of the CMB lensing convergence and cosmic shear convergence. The shaded region denotes the $1\sigma$ error bar estimated from the Gaussian covariance with $f_{\rm sky}=0.45$, as described in Sec.~\ref{Sec:forecast_setup}. The CMB lensing results (orange and red) include lensing reconstruction noise for Simons Observatory and CMB-S4, while the shear results (blue) include shape noise and assume a single tomographic bin; the noise power spectra are shown in dash-dotted. The black dotted curves indicate linear-theory predictions. The cosmic shear power spectrum is significantly more sensitive to non-linear structure formation than the CMB lensing power spectrum, but also has much higher signal-to-noise at small scales (large $\ell$). The power spectra are computed using FFTLog for $\ell<50$ and using the Limber approximation for $\ell\geq 50.$ \textbf{Right}: Assumed redshift source distribution for the cosmic shear forecasts presented in this work based on an LSST/{\it Euclid}-like survey. The full sample (solid) is normalized to unit integral and the three tomographic bins (dashed) are normalized to $1/3$ each. For comparison, we include the CMB lensing kernel $W^{\kappa_{\rm CMB}}(z)/H(z)$ computed using Eq.~\eqref{eq:CMB_lensing_window} with an arbitrary normalization. } \label{fig:Cl_CMB_and_shear}
\end{figure*}

\subsection{Survey specifications}\label{Subsec:survey_specs}
\subsubsection{CMB lensing experiments}

For the CMB lensing analysis, we consider two CMB experiments: Simons Observatory~\cite{SimonsObservatory:2018koc} and CMB-S4~\cite{CMB-S4:2016ple,Abazajian:2019eic}. We assume a fractional sky coverage $f_{\rm sky}=0.45$ for both surveys, for simplicity. The observed angular power spectrum of the CMB lensing convergence, used to compute the bispectrum covariance in Eq.~\eqref{eq:reduced_bspec_cov}, is 
\begin{align}
    \tilde{C}_{\ell}=C_{\ell}^{\kappa_{\rm CMB}}+N_{\ell} \,,
\end{align}
where $N_{\ell}$ is the CMB lensing reconstruction noise. Here, we assume that the CMB lensing convergence is reconstructed using an iterative estimator~\cite{Hirata:2003ka, Carron:2017mqf} and model the reconstruction noise using the iterative noise curves from~\cite{Robertson:2023xkg}. The convergence power spectrum $C_{\ell}^{\kappa_{\rm CMB}}$ is computed using Eqs.~\eqref{eq:angular_power_spec_full} and~\eqref{eq:Cl_limber} where we compute the matter power spectrum using \textsc{halofit}~\cite{Takahashi:2012em} to model contributions from non-linear structure formation.\footnote{We use the linear power spectrum for the soft mode when computing the non-Limber integrals since the FFTLog algorithm requires that the time dependence of the integrand factorizes. This has negligible impact on our results because these scales are well described by linear theory. } 

The left panel of Fig.~\ref{fig:Cl_CMB_and_shear} shows the CMB lensing convergence power spectrum and its covariance, including the cosmic variance and reconstruction noise expected for Simons Observatory and CMB-S4. The CMB lensing convergence power spectrum is cosmic-variance-limited (on the observed sky fraction) up to $\ell\simeq 350$ (700) for Simons Observatory (CMB-S4).

\subsubsection{Cosmic shear experiments}\label{Subsec:cosmic_shear_experiments}

For the cosmic shear analysis, we consider a generic Stage IV photometric galaxy survey with specifications similar to those expected of LSST~\cite{LSSTScience:2009jmu} and {\it Euclid}~\cite{EUCLID:2011zbd}.  As for the CMB, we assume that the cosmic shear surveys have a fractional sky coverage $f_{\rm sky}=0.45$. When cross-correlating cosmic shear and CMB lensing, we assume the surveys fully overlap. 

We model the \emph{true} source galaxy redshift distribution as
\begin{equation}
    p_s^{\rm true}(z)\propto z^2\exp\left[-(z/z_0)^\alpha\right]\,,
\end{equation}
with $z_0=0.28$ and $\alpha=0.9$ as specified in the LSST Dark Energy Science Collaboration Science Requirements Document~\cite{LSSTDarkEnergyScience:2018jkl}. To account for photometric-redshift uncertainties, we assume that the probability distribution for the observed photometric redshift, $z_{\rm ph}$, given the true galaxy redshift, $z$, follows a Gaussian distribution,
\begin{equation}
    \mathcal{P}(z_{\rm ph}|z)=\frac{1}{\sqrt{2\pi\sigma^2(z)}}\exp{\left[-\frac{1}{2}\left(\frac{z-z_{\rm ph}}{\sigma^2(z)} \right)^2\right]} \,,
\end{equation}
with uncertainty $\sigma(z)=0.05(1+z)$~\cite{LSSTDarkEnergyScience:2018jkl}. The source distribution in a tomographic bin $z^{(i)}_{\rm min}\leq z\leq z^{(i)}_{\rm max}$ is then 
\begin{equation}
    p_s^{(i)}(z)=p_s^{\rm true}(z)\int_{z^{(i)}_{\rm min}}^{ z^{(i)}_{\rm max}}\, dz^\prime\,\mathcal{P}\left(z^\prime|z\right) \,.
\end{equation}
For our fiducial forecast, we divide the source distribution into three tomographic bins with an equal number of galaxies in each bin. We investigate the impact of varying the number of tomographic bins on our parameter constraints in Sec.~\ref{Sec:shear_results}. 

The observed angular power spectrum of the cosmic shear convergence used to compute the bispectrum covariance in Eq.~\eqref{eq:reduced_bspec_cov} is 
\begin{equation}
    \tilde{C}_{\ell}^{(ij)}=C_{\ell}^{\kappa_g,(ij)}+\delta_{ij}^K\frac{\sigma_\epsilon^2}{\bar{n}_g^{(i)}} \,,
\end{equation}
where we have included the shape noise contribution arising from the intrinsic ellipticities of galaxies. We set the effective number density of galaxies to $\bar{n}_g=31~{\rm arcmin}^{-2}$ and assume an intrinsic rms ellipticity $\sigma_\epsilon=0.26$~\cite{Chang:2013xja}. When dividing the galaxy source sample into $N_{\rm tomo}$ tomographic bins, the effective number density is $\bar{n}_g^{(i)}=\bar{n}_g/N_{\rm tomo}$.

 The left panel of Fig.~\ref{fig:Cl_CMB_and_shear} includes the cosmic shear convergence power spectrum assuming $N_{\rm tomo}=1$ as well as the noise contribution due to shape noise. The cosmic shear power spectrum is cosmic-variance-limited (on the observed sky fraction) up to $\ell\simeq 1800$ assuming $N_{\rm tomo}=1$ and the survey specifications used in this work. The right panel of Fig.~\ref{fig:Cl_CMB_and_shear} shows the assumed galaxy redshift source distribution in a single tomographic bin and in three tomographic bins, as well as the CMB lensing kernel.

\subsection{Fisher Matrix}\label{sec:forecast_setup_fisher_mat}

To estimate how well lensing convergence bispectra can constrain $f_{\rm NL}$, we adopt the Fisher matrix formalism. For our fiducial forecasts, we assume that the auto- and cross-bispectra of the convergence fields are the only observables and that $f_{\rm NL}$ is the only free parameter.\footnote{Note that this differs from the analysis in Ref.~\cite{Goldstein:2022hgr}, which used a joint likelihood in the 3D matter power spectrum and bispectrum to take advantage of the sample variance cancellation associated with the significant correlation between the squeezed bispectrum $B(\vq,\vk)$ and the soft mode power spectrum $P(q)$. It would be interesting to consider whether a similar cancellation applies to the angular power spectrum $C_{\ell_1}$ and the squeezed angular bispectrum $b_{\ell_1\ell_2\ell_3}$; however, we leave this to future work because it would require including the non-Gaussian contributions to the power spectrum and bispectrum covariances, since the power spectrum and bispectrum are uncorrelated in the Gaussian limit. Additionally, $b_{\ell_1\ell_2\ell_3}$ and $C_{\ell_1}$ have different projection integrals, which  likely reduces their correlation.} We also assume a Gaussian likelihood for the observed convergence bispectra. The Fisher matrix is then given by 
\begin{equation}\label{eq:1x1_fisher_matrix}
    \mathcal{F}=\sum\limits_{\substack{ijk\\ abc}}\sum_{\ell_1\ell_2\ell_3}\frac{\partial\,b^{(ijk)}_{\ell_1\ell_2\ell_3}}{\partial f_{\rm NL}}{\rm{Cov}}^{-1}\left[b^{(ijk)}_{\ell_1\ell_2\ell_3},b^{(abc)}_{\ell_1\ell_2\ell_3}\right]\frac{\partial\,b^{(abc)}_{\ell_1\ell_2\ell_3}}{\partial f_{\rm NL}} \,,
\end{equation}
where $\partial\,b^{(ijk)}_{\ell_1\ell_2\ell_3}/\partial f_{\rm NL}$ is computed using Eqs.~\eqref{eq: sq_Bk_fNL_3Dmatter},~\eqref{eq:reduced_bispec_squeezed}, and~\eqref{eq:reduced_bispec_limber}, and the covariance is computed using Eq.~\eqref{eq:reduced_bspec_cov}. The Cram{\'e}r--Rao bound guarantees that the minimum variance of an unbiased estimator of a given parameter is equal to the inverse Fisher matrix element of the associated parameter, hence $\sigma_{f_{\rm NL}}\geq 1/\sqrt{\mathcal{F}}.$

The sum over $i,j,k$ and $a,b,c$ in Eq.~\eqref{eq:1x1_fisher_matrix} runs over all tomographic bins included in the analysis. For example,  when considering CMB lensing and cosmic shear cross-correlations, the $i,j,k$ and $a,b,c$ range from 1 to $N_{\rm tomo}+1$. The sum over $\ell_1,$ $\ell_2,$ and $\ell_3$ runs over all possible multipoles satisfying the following criteria:
\begin{itemize}
    \item $\ell_{\rm min}^{\rm soft}\leq \ell_1\leq\ell_{\rm max}^{\rm soft}\,$;
    \item $\ell_{\rm min}^{\rm hard}\leq \ell_2\leq\ell_{\rm max}^{\rm hard}\,$;
    \item $|\ell_i-\ell_j|\leq \ell_k\leq \ell_i+\ell_j\,$;
    \item $\ell_1+\ell_2+\ell_3=\rm{even}\,$.
\end{itemize}
The first two criteria ensure that the triangles are sufficiently squeezed such that the bispectrum model in Eq.~\eqref{eq: sq_Bk_fNL_3Dmatter} applies.\footnote{In principle, different scale cuts could be imposed for each tomographic bin, as well as for CMB lensing versus cosmic shear, because these measurements probe different physical scales and redshifts and are sensitive to different systematics. We do not account for this in our forecasts because quantifying the exact range of multipoles for which our bispectrum model applies for a given convergence field would require simulations. Nevertheless, we note that this approach of varying scale cuts could be useful in practice.} The last two criteria arise from momentum conservation and parity, respectively. For our forecasts, we fix the maximum soft multipole $\ell_{\rm max}^{\rm soft}=100$ and the minimum hard multipole $\ell_{\rm min}^{\rm hard}=150$.\footnote{This choice of multipoles is somewhat optimistic and a precise verification of the range of scales for which our model is valid would require analyzing simulations, similar to what was done for the 3D matter bispectrum in~\cite{Goldstein:2022hgr}. Nevertheless, since most of the $f_{\rm NL}$ information is coming from the lowest $\ell_1$ bins, our results are largely insensitive to the precise choice of $\ell_{\rm max}^{\rm soft}$. } In theory, the minimum soft multipole $\ell_{\rm min}^{\rm soft}$ is determined by the largest scales observed in the survey; in practice, however, the largest scales can be plagued by observational and theoretical systematics, including foreground contamination (note that high-$\ell$ CMB temperature and polarization foregrounds lead to low-$\ell$ biases in the reconstructed lensing convergence field)~\cite{Osborne:2013nna,vanEngelen:2013rla,Ferraro:2017fac,Cai:2021hnb}, relativistic corrections~\cite{Bernardeau:2011tc}, and non-Gaussianity of the likelihood~\cite{Wang:2018xuy, Tucci:2023bag}. Consequently, we will consider $\ell_1^{\rm min}=2$, 10, and 20. We vary the maximum hard multipole $\ell_{\rm max}^{\rm hard}$ to determine the scales at which our constraints saturate due to noise.

In addition to the forecasts assuming $f_{\rm NL}$ is the only free parameter, we present forecasts where we marginalize over the leading-order gravitational contribution to the matter bispectrum, $a_0(k, \chi_k)$ (see Eq. \ref{eq:b_3D_grav}). Directly marginalizing over $a_0(k, \chi_k)$ is challenging because $a_0(k, \chi_k)$ depends on scale and redshift, both of which enter the bispectrum evaluations in Eqs.~\eqref{eq:reduced_bispec_squeezed} and \eqref{eq:reduced_bispec_limber}. Therefore, we assume that $a_0(k, \chi_k)$ varies slowly over the integration volume and can thus be approximated by a single coefficient $\bar{a}^{(i)}_0$, where the index $(i)$ is used to indicate that this parameter depends on the redshift kernel of the bispectrum considered.\footnote{We find that approximating $a_0(k,\chi_k)$ as a constant evaluated at the peak of the lensing kernel can bias the bispectrum by at most $\simeq 10\%$ compared to full numerical integration using Eq.~\eqref{eq:a0_CR}.} For simplicity, we present these marginalized forecasts using only a single convergence field, hence the only free parameters are $\bar{a}_0$ and $f_{\rm NL}$. The $2\times 2$ Fisher matrix is
\begin{equation}\label{eq:2x2_fisher_matrix}
    \mathcal{F}_{\alpha\beta}=\sum_{\ell_1\ell_2\ell_3}\frac{\partial\,b_{\ell_1\ell_2\ell_3}}{\partial\theta_\alpha}{\rm{Cov}}^{-1}\left[b_{\ell_1\ell_2\ell_3},b_{\ell_1\ell_2\ell_3}\right]\frac{\partial\,b_{\ell_1\ell_2\ell_3}}{\partial \theta_\beta} \,,
\end{equation}
where $\theta_\alpha\in \{ f_{\rm NL}, \bar{a}_0\}.$ The Fisher error on $f_{\rm NL}$ is then $\sigma_{f_{\rm NL}}=\sqrt{\mathcal{F}_{f_{\rm NL}f_{\rm NL}}^{-1}}$.

Finally, since the scale cuts used in this work typically contain $\mathcal{O}(10^6)$ triangles, we adopt a binning strategy based on Ref.~\cite{Takada:2003ef}. In particular, we bin the first two multipoles with bin widths $\Delta\ell_1=5$ and $\Delta\ell_2=15$ and rescale the covariance by a factor of $(\Delta\ell_1\Delta\ell_2)^{-1}$. We let $\ell_3$ range over all available multipoles subject to the triangle inequality, the parity constraint, and the ordering $\ell_2\leq \ell_3$. We have verified that this binning approximation has negligible impact on our forecasts.

\begin{figure*}[!t]
\includegraphics[width=\linewidth]{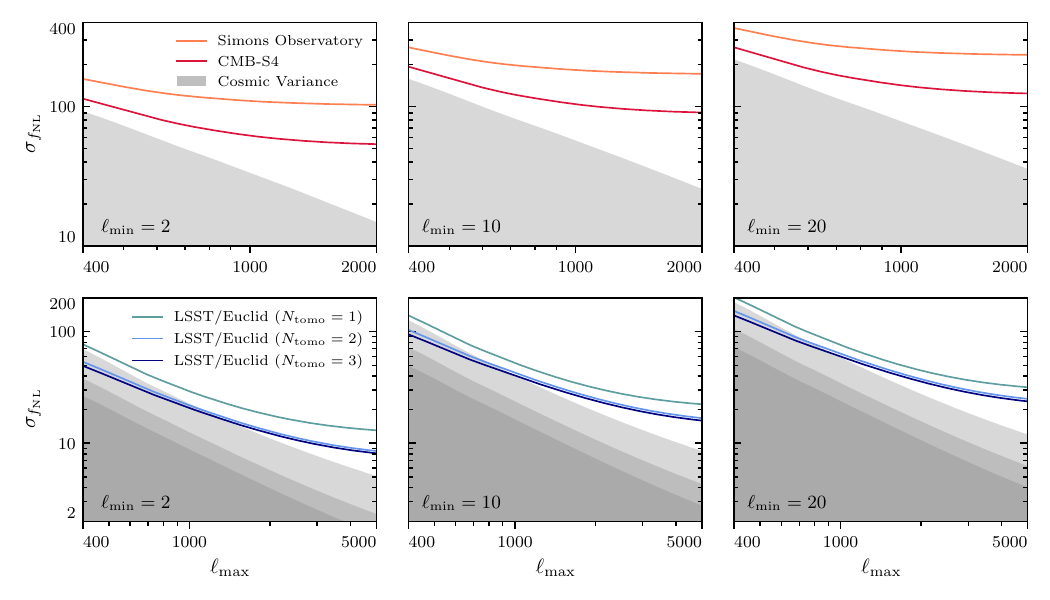}
\caption{Forecasted error on $f_{\rm NL}$ from the non-linear squeezed bispectrum of CMB lensing (\textbf{top}) and cosmic shear (\textbf{bottom}) using a non-perturbative bispectrum model based on the cosmological consistency relations that is independent of the non-Gaussian bias, $b_\phi$. The error is shown a function of the maximum multipole $\ell_{\rm max}$ for several values of the minimum multipole $\ell_{\rm min}=2, \ 10, $ and $20$. The forecasts are sensitive to the minimum multipole $\ell_{\rm min}.$ Whereas the CMB lensing forecasts saturate around $\ell_{\rm max}=1000$ due to lensing reconstruction noise, the shear forecasts begin to plateau around $\ell_{\rm max}=3000$ due to shape noise. The cosmic shear forecasts improve if we divide the source galaxy sample into two tomographic bins, but there is little to gain with $N_{\rm tomo}\geq 3$. The cosmic variance error is shown by the grey bands. The three grey bands in the shear forecasts correspond to the number of tomographic bins. All forecasts assume fractional sky coverage $f_{\rm sky}=0.45$.} \label{fig:forecast_CMB_shear_alone}
\end{figure*}

\section{Results}\label{Sec:results}
\subsection{CMB Lensing}

The top panels of Fig.~\ref{fig:forecast_CMB_shear_alone} show the expected error on $f_{\rm NL}$ from the CMB lensing bispectrum as a function of the maximum hard multipole for three different choices of the minimum soft multipole. The shaded region denotes the cosmic variance error, which assumes perfect knowledge of the lensing potential ({i.e.}, $N_\ell = 0$).  The forecasts for Simons Observatory (CMB-S4) saturate by $\ell_{\rm max}\simeq 800~(1200).$ For the fiducial Simons Observatory analysis with $\ell_{\rm min}=10$ and $\ell_{\rm max}=1400$, the error on $f_{\rm NL}$ is $\sigma_{f_{\rm NL}}=175$; with these scale cuts, CMB-S4 can improve the constraint by almost a factor of two, yielding $\sigma_{f_{\rm NL}}=95$. Notably, the forecasted error on $f_{\rm NL}$ is very sensitive to the lowest multipole. Assuming the CMB lensing bispectrum can be reliably reconstructed down to $\ell_{\rm min}=2$, then the method presented here constrains $f_{\rm NL}$ with precision $\sigma_{f_{\rm NL}}=105~(56)$ for Simons Observatory (CMB-S4).

The findings of this section indicate that forthcoming measurements of the CMB lensing bispectrum from Simons Observatory and CMB-S4 may not offer competitive constraints on $f_{\rm NL}$ compared with those obtained from the primary CMB or from LSS constraints based on the scale-dependent bias, though the constraints are on different characteristic scales. Nevertheless, the forecasted error on $f_{\rm NL}$ is still better than the tightest current $b_\phi$--independent LSS constraints on $f_{\rm NL}$~\cite{Cabass:2022ymb}. The main limitation of these forecasts is the CMB lensing reconstruction noise, which severely restricts our ability to precisely measure highly squeezed configurations of the CMB lensing bispectrum. In the more long-term future, the method presented here could prove to be quite powerful for a CMB experiment that is cosmic-variance-limited up to much smaller scales, such as CMB-HD~\cite{CMB-HD:2022bsz}, although one should properly include the non-Gaussian covariance to properly assess how much information is practically available at smaller scales.

\begin{figure*}[!t]
\includegraphics[width=0.99\linewidth]{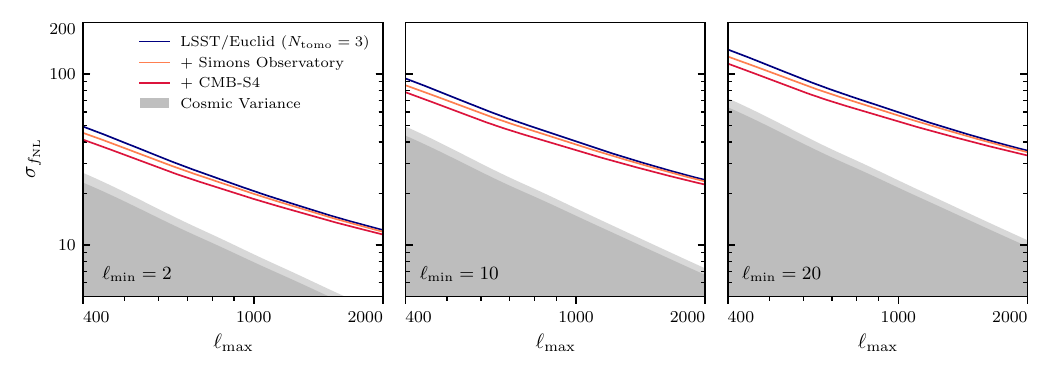}
\caption{As Fig.~\ref{fig:forecast_CMB_shear_alone}, but showing forecasts for a joint analysis of the CMB lensing and cosmic shear auto- and cross-bispectra assuming three tomographic redshift bins. At low maximum multipoles ($\ell_{\rm max}\approx400$), including CMB lensing convergence measurements expected from Simons Observatory (CMB-S4) improves the forecasted error on $f_{\rm NL}$ by $\simeq 10\%$ (20\%). At larger maximum multipoles ($\ell_{\rm max}\approx2000)$, the constraint on $f_{\rm NL}$ is dominated by the shear bispectrum with improvements of only $\simeq 1$\% (5\%) when including the Simons Observatory (CMB-S4) lensing convergence. The two grey bands correspond to the shear and shear$\times$CMB lensing cosmic variance errors. } \label{fig:forecast_CMB_cross_shear}
\end{figure*}

\subsection{Cosmic Shear}\label{Sec:shear_results}

The bottom panels of Fig.~\ref{fig:forecast_CMB_shear_alone} show the expected error on $f_{\rm NL}$ from cosmic shear bispectra as a function of $\ell_{\rm max}$. Interestingly, even for a single tomographic bin and fixed scale cuts, the cosmic variance error on $f_{\rm NL}$ from the shear bispectrum is approximately $20-40\%$ smaller than that from the CMB lensing bispectrum. This is because the squeezed limit convergence bispectrum is proportional to an integral over $\partial P(k)/\partial \log(\sigma_8^2)$, whereas the noise is proportional to $P(k)$. As a result, the signal-to-noise is enhanced by the logarithmic derivative $\partial \log P(k)/\partial \log(\sigma_8^2)$, which is more pronounced at low redshifts and small scales.\footnote{For the scales and redshifts considered here, the logarithmic derivative is greater than one; however, the situation can reverse at very small scales and low redshifts (see Fig.~\ref{fig:log_deriv} in the Appendix).} This is discussed in more detail in Appendix~\ref{appendix:log_deriv}.

The forecasted constraints on $f_{\rm NL}$ from the shear bispectrum measured from an LSST/{\it Euclid}-like survey saturate at much higher $\ell_{\rm max}$ than the forecasts from the CMB lensing bispectrum measured by Simons Observatory or CMB-S4. Assuming a single tomographic bin with $\ell_{\rm min}=10$ and $\ell_{\rm max}=5000$, the shear bispectrum constrains $f_{\rm NL}$ with precision $\sigma_{f_{\rm NL}}=22$. The forecasted error improves to $\sigma_{f_{\rm NL}}=16$ using three tomographic bins, indicating that tomographic information of the source galaxies can significantly improve constraints on $f_{\rm NL}$. Similar to the findings of Ref.~\cite{Takada:2003ef}, our forecasts do not improve significantly if we include more than three tomographic bins. Taken at face value, the results of this section suggest that upcoming measurements of tomographic lensing bispectra from Stage-IV shear surveys provide a more promising avenue towards constraining $f_{\rm NL}$ than upcoming measurements of the CMB lensing bispectrum, and one that could be vital in confirming detections from the scale-dependent bias method, such as \citep{Rezaie:2023lvi}. However, it is important to note that, since the shear bispectrum probes smaller scales and lower redshifts than the CMB lensing bispectrum, the shear forecasts  would likely be more sensitive to the non-Gaussian covariance, which we have ignored.

\subsection{CMB Lensing $\times$ Cosmic Shear}\label{subsec:lens_shear}

Fig.~\ref{fig:forecast_CMB_cross_shear} shows the forecasted error on $f_{\rm NL}$ for a joint analysis of the CMB lensing convergence and the cosmic shear auto- and cross-bispectra. These forecasts assume that the shear field is measured in three tomographic redshift bins. Assuming $\ell_{\rm max}=400$ and $\ell_{\rm min}=10$, including CMB lensing measurements from Simons Observatory (CMB-S4) reduces the error on $f_{\rm NL}$ from 94 to 86 (78). The improvement from a joint analysis is much less significant at smaller scales, where the signal-to-noise of the shear bispectrum is significantly larger than that of the CMB lensing convergence bispectrum. Indeed, assuming $\ell_{\rm max}=2000$ and $\ell_{\rm min}=10$, a joint analysis of CMB-S4 and an LSST/{\it Euclid}-like shear experiment leads to a less than 5\% improvement on the constraint on $f_{\rm NL}$ compared to a shear-only analysis.

Although these results suggest that there is little to gain from a joint analysis of CMB lensing and cosmic shear in comparison to a shear-only analysis, the situation may differ considerably in practice. For instance, a realistic analysis would likely impose different scale cuts for the CMB lensing measurements and the cosmic shear measurements to account for their distinct systematics (with CMB lensing potentially providing easier access to low $\ell$, modulo foreground biases or other reconstruction systematics). If one uses only the CMB lensing convergence to measure large-scale modes, then the joint analysis is restricted to 16 of the 64 total bispectra combinations, assuming $N_{\rm tomo}=3$. Under these conditions, with $\ell_{\rm min}=10$ and $\ell_{\rm max}=2000$, the resulting error on $f_{\rm NL}$ is $\sigma_{f_{\rm NL}}=32$ for CMB-S4. Whereas this value is worse than the forecasted $\sigma_{f_{\rm NL}}=23$ from a joint analysis of all 64 bispectra combinations, it still represents a significant improvement over the $\sigma_{f_{\rm NL}}=95$ value obtained for a CMB-S4 lensing convergence bispectrum-only analysis with these scale cuts.

 \begin{figure}[!t]
    \centering
    \includegraphics[width=0.95\linewidth]{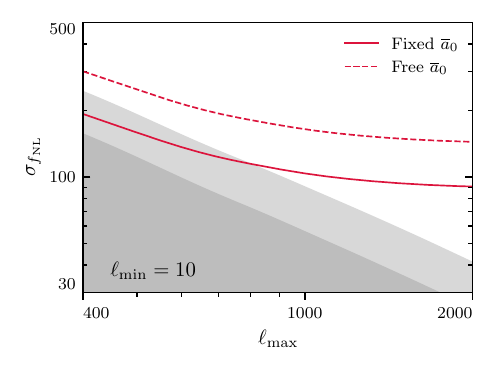}
    \caption{Forecasted error on $f_{\rm NL}$ from the CMB lensing bispectrum measured by CMB-S4 with and without marginalization over gravitational non-Gaussianity. The solid result assumes perfect knowledge of the leading-order gravitational non-Gaussianity parameter, $a_0(k,\chi_k)$. The dashed line marginalizes over $a_0(k,\chi_k)$ using the procedure described in Sec.~\ref{sec:forecast_setup_fisher_mat}. Marginalization over $a_0(k,\chi_k)$ weakens the constraint on $f_{\rm NL}$ by approximately 50\%. The two grey shaded regions correspond to the cosmic-variance-limited error on $f_{\rm NL}$ from the CMB lensing bispectrum with and without marginalizing over $a_0(k,\chi_k)$. }
    \label{fig:grav_non_gauss_marge}
\end{figure}

\subsection{Marginalizing over gravitational non-Gaussianity}\label{Sec:marginalization_grav_non_gauss}

Finally, we consider how our forecasts would change if we no longer assume perfect knowledge of the leading-order gravitational non-Gaussianity contribution,  $a_0(k, \chi_k)$ (see Eq. \ref{eq:b_3D_grav}). The left panel of Fig.~\ref{fig:grav_non_gauss_marge} compares the forecasted error on $f_{\rm NL}$ from the CMB lensing bispectrum with and without marginalization over the leading-order gravitational non-Gaussianity parameter $a_0(k,\chi_k)$ using the marginalization procedure described in Sec.~\ref{sec:forecast_setup_fisher_mat}. The red curves assume CMB-S4 lensing reconstruction noise and the grey shaded regions denote the cosmic variance error. Marginalizing over gravitational non-Gaussianty degrades the constraining power quite significantly, with the forecasted error on $f_{\rm NL}$ increasing from $\sigma_{f_{\rm NL}}=95$ to $150$, assuming $\ell_{\rm max}=1400$. This loss in constraining power arises from the large correlation between the primordial contribution to the matter bispectrum, Eq.~\eqref{eq: sq_Bk_fNL_3Dmatter}, and the leading-order gravitational contribution, Eq.~\eqref{eq:b_3D_grav}, which, at equal times, is only broken by the $1/(q^2T(q))$ term.\footnote{The $k$-dependence of the $a_0(k,\chi_k)$ coefficient also breaks this correlation; however, this $k$-dependence is small and is neglected in our forecasts.} These findings are in agreement with Ref.~\cite{Goldstein:2022hgr}, which found a sizeable cross-correlation $(r\simeq 0.7$ at $z=0.97$) between $\bar{a}_0$ and $f_{\rm NL}.$

Based on these results, it is clear that marginalizing over the gravitational contribution to the squeezed matter bispectrum significantly degrades constraints on $f_{\rm NL}$ using the method presented here. Nevertheless, there are several possible ways to extend our analysis that could improve constraints on $a_0(k,\chi_k)$ and, hence, $f_{\rm NL}$. First of all, since $a_0(k,\chi_k)\simeq B(q,k)/P(q)P(k)$, a joint analysis of the convergence bispectrum and convergence power spectrum could improve constraints on $a_0(k,\chi_k)$. Ref.~\cite{Goldstein:2022hgr} exploited this sample variance cancellation in 3D and found that it significantly improved constraints on $f_{\rm NL}$. A precise determination of the impact of sample variance cancellation is beyond the scope of this work because it requires non-Gaussian covariances. Secondly, cross-correlation of CMB lensing and cosmic shear maps could help constrain $a_0(k,\chi_k)$, since both the CMB lensing bispectrum and the cosmic shear bispectrum depend on differently weighted integrals over $a_0(k,\chi_k).$

\section{Conclusions}\label{Sec:conclusions}

Constraining the amplitude of local primordial non-Gaussianity, $f_{\rm NL}$, is a key goal of upcoming cosmological surveys. Here, we have quantified how well squeezed configurations of the CMB lensing convergence and cosmic shear bispectra, as measured by forthcoming experiments, can constrain $f_{\rm NL}$. Our method has utilized a non-perturbative model for the squeezed bispectrum based on the cosmological consistency relations, which is, in principle, applicable down to very small scales (high $\ell_{\rm max}$). In practice, we have found that, even with a joint analysis of the auto- and cross-bispectra of convergence measurements from an LSST/{\it Euclid}-like experiment and CMB-S4, it will be difficult to constrain $\sigma_{f_{\rm NL}}\lesssim 10$.

In the modern world, with an abundance of forecasts predicting $\sigma_{f_{\rm NL}}\lesssim 1$ using Stage-IV surveys (with varying degrees of optimism in their assumptions), it is natural to question the utility of the approach presented here. We emphasize that the primary advantage of using the squeezed lensing bispectrum to constrain $f_{\rm NL}$ is that it actually constrains $f_{\rm NL}.$ This stands in contrast to most existing methods for constraining $f_{\rm NL}$ using LSS, which rely on the scale-dependent bias, and hence require accurate knowledge of galaxy formation physics to obtain a direct constraint on $f_{\rm NL}.$\footnote{In a not entirely unrealistic, but still unrealistic, scenario in which cosmologists find themselves debating $b_\phi f_{\rm NL}$ vs. $f_{\rm NL}$, the method outlined here could potentially quell the controversy.} Moreover, the squeezed lensing bispectrum is sensitive to different large-scale systematics compared to alternative methods to constrain $f_{\rm NL}$ (e.g., the galaxy power spectrum); thus, our method can be used as a valuable cross-check for future constraints on $f_{\rm NL}$ using LSS data. Finally, our approach could be extended to include galaxy clustering information. Combining galaxy clustering with CMB lensing and/or cosmic shear has already shown great promise as a probe of $f_{\rm NL}$ using two-point statistics~\cite{Schmittfull:2017ffw}; therefore, a joint analysis of galaxy clustering, cosmic shear, and CMB lensing two-point and three-point functions could offer highly informative results. Although these constraints would no longer be independent of $b_\phi$, the potential for sample variance cancellations and forming cross-bispectra with the galaxy density field as the long mode (similar to the discussion in Sec.~\ref{subsec:lens_shear} for CMB lensing) makes this approach particularly promising from a modeling perspective.
It is also worth noting that aside from constraining constraining $f_{\rm NL}$, checking the consistency relation with data constitutes a test of equivalence principle as well. 
For that purpose, a measurement of the bispectrum involving different galaxy populations would be needed \cite{Creminelli:2013nua}.

As with all forecasts, our results depend on several modeling assumptions. First of all, we have ignored various systematics that could impact the CMB lensing and/or cosmic shear bispectra, such as post-Born corrections~\cite{Pratten:2016dsm, Fabbian:2017wfp}, reconstruction noise biases~\cite{Kalaja:2022xhi}, intrinsic alignments~\cite{Troxel:2014dba}, source clustering~\cite{DES:2023ycm}, and relativistic effects~\cite{Bernardeau:2011tc}. Furthermore, we omitted baryonic effects in our analysis, despite their significant impact on the correlators of weak lensing convergence across various scales~\cite{Ferlito:2023gum}. One may worry that baryons pose a significant challenge to the methodology presented here, which uses measurements from extremely small scales. However, it is important to note that, since the cosmological consistency relations are still satisfied with baryons, our model that marginalizes over gravitational non-Gaussianity remains valid as long as baryonic corrections are included when computing the response function, $\partial \log(P(k, z))/\partial\log(\sigma_8^2)$. Previous studies have shown that these corrections are relatively small~\cite{Barreira:2019ckp}.\footnote{
It is worth reiterating that gravitational non-Gaussianity cannot produce poles in the squeezed bispectrum to soft power spectrum ratio, regardless of the complexity of the baryonic feedback effects \cite{Esposito:2019jkb,Horn:2014rta}. Thus observing such poles immediately points to PNG (beyond that expected from single-field slow roll inflation) or equivalence principle violation. To infer parameters such as $f_{\rm NL}$ from the residue of such poles though, does require incorporating baryonic corrections in the response function.}
Finally, to obtain an upper bound on our method's utility, we have neglected non-Gaussian contributions to the bispectrum covariance. These contributions could significantly degrade the cosmic shear forecasts at high $\ell_{\rm max}$, but we leave this to a future work.

There are several ways to build upon our analysis. An immediate follow-up would be to use simulations to explicitly verify the range of scale cuts for which our bispectrum model is valid and to quantify the impact of the non-Gaussian covariance. Such an analysis could be compared with the results of Ref.~\cite{Anbajagane:2023wif}, which assessed the information content of primordial non-Gaussianity in the lensing convergence field at non-linear scales. Additionally, it would be valuable to generalize the method presented here to include galaxy clustering statistics and information from the scale-dependent bias. Finally, the method presented in this work can be readily extended to test alternative non-standard cosmological scenarios, such as quasi-single field inflation~\cite{Assassi:2012zq, Chen:2009zp, Baumann:2010tm, Noumi:2012vr} or equivalence-principle-violating physics~\cite{Creminelli:2013nua}, all of which generate analogous poles in the squeezed bispectrum. An observational test of the LSS consistency relations using weak lensing bispectra would also complement the recent test of the cosmological consistency relations using the anisotropic three-point correlation function~\cite{Sugiyama:2023zvd}.

\acknowledgments
{\footnotesize
 We acknowledge computing resources from Columbia University's Shared Research Computing Facility project, which is supported by NIH Research Facility Improvement Grant 1G20RR030893-01, and associated funds from the New York State Empire State Development, Division of Science Technology and Innovation (NYSTAR) Contract C090171, both awarded April 15, 2010. We acknowledge the Texas Advanced Computing Center (TACC) at The University of Texas at Austin for providing HPC resources that have contributed to the research results reported within this paper. OHEP is a Junior Fellow of the Simons Society of Fellows and edited this draft at a speed inspired by the Costa Rican sloth population. JCH acknowledges support from
NSF grant AST-2108536, NASA grants 21-ATP21-0129 and 22-
ADAP22-0145, the Sloan Foundation, and the Simons Foundation. LH acknowledges support by the DOE DE-SC011941 and a Simons Fellowship in Theoretical Physics.
} 

\bibliographystyle{apsrev4-1}
\bibliography{biblio}

\clearpage
\appendix
\onecolumngrid
\begin{figure*}[!t]
    \centering
    \includegraphics[width=0.99\linewidth]{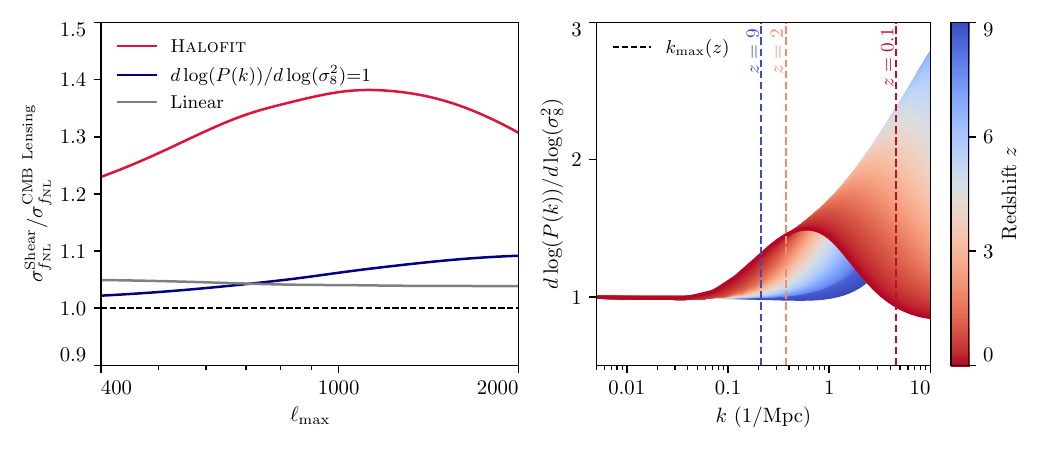}
    \caption{\textbf{Left}: Ratio of cosmic variance error on $f_{\rm NL}$ as a function of $\ell_{\rm max}$ from the cosmic shear bispectrum with $N_{\rm tomo}=1$ to that from the CMB lensing convergence bispectrum for a variety of non-linear modeling choices. The results from the forecasts of the main text are shown in red where the shear bispectrum yields $\sim30\%$ tighter constraints on $f_{\rm NL}$ than the CMB lensing bispectrum. If we neglect the non-linear enhancement from the logarithmic derivative (blue), then the forecasted error on $f_{\rm NL}$ agrees within 10\% between the two approaches. Finally, if we use linear theory, then the two constraints are consistent to within 4\%. The slight discrepancy arises from differences in the projection kernels. \textbf{Right}: \textsc{halofit}  prediction for $\partial \log(P(k, z))/\partial\log(\sigma_8^2)$ over redshifts $0<z<9.$ Vertical lines indicate the approximate maximum 3D wavenumber $k_{\rm max}\simeq \ell_{\rm max}/{\chi(z)}$ probed at a certain redshift assuming $\ell_{\rm max}=2000.$ For $z\gtrsim9$ the \textsc{halofit} predictions can become unstable so we fix the logarithmic derivative to $1$ ({i.e.}, linear theory) which is accurate at all scales of interest. }
    \label{fig:log_deriv}
\end{figure*}
\twocolumngrid
\section{Impact of non-linearities on cosmic shear and CMB lensing forecasts}\label{appendix:log_deriv}

In this section, we discuss the impact of non-linear structure formation on the forecast results in the main text. In the realm of 3D matter distributions, and ignoring gravitational non-linearities, the redshift dependence of the bispectrum scales as $B\sim P^2/D(z)\sim D^3(z)$, whereas the redshift dependence of the covariance scales as $P^3\approx D^6$. Consequently, the signal-to-noise should be roughly independent of redshift. Nevertheless, our forecast results show that the cosmic variance error on $f_{\rm NL}$ can differ by up to 50\% between the cosmic shear forecasts and the CMB lensing forecasts for fixed scale cuts and assuming a single tomographic bin. We investigate the source of this discrepancy in this section.

The right panel of Fig.~\ref{fig:log_deriv} compares the ratio of the cosmic variance error on $f_{\rm NL}$ as a function of $\ell_{\rm max}$ for a cosmic shear analysis with a single tomographic bin compared to that from a CMB lensing analysis for a variety of non-linear modeling choices. These results assume the fiducial scale cuts in the main text, with $\ell_{\rm min}=10$. The red line corresponds to the analysis choices used in the main forecasts, where all power spectra and the logarithmic derivative $\partial \log(P(k))/\partial \log(\sigma_8^2)$ are computed using \textsc{halofit}. In this case, the shear bispectrum provides  $\sim30\%$ tighter constraints on $f_{\rm NL}$ than the CMB lensing bispectrum. However, if we fix the logarithmic derivative to the linear theory prediction of unity (blue), then the improvement from cosmic shear diminishes considerably. This shows that the main source of discrepancy between the cosmic variance CMB lensing and cosmic shear forecasts is due to the non-perturbative enhancement of the squeezed matter bispectrum due to local PNG. It remains to be seen to what extent this also reduces our constraining power due to the associated non-Gaussian covariance. Finally, we can use linear theory to also compute the bispectrum Eq.~\eqref{eq: sq_Bk_fNL_3Dmatter} and its covariance (grey). In this case, the cosmic shear and CMB lensing forecasts are consistent to within $\simeq 4\%$, with the residual attributed to differences in the projection kernels.

The right panel of Fig.~\ref{fig:log_deriv} shows the \textsc{halofit} prediction for the logarithmic derivative $\partial \log(P(k))/\partial \log(\sigma_8^2)$ for a range of redshifts. At high redshifts and large wavenumbers the logarithmic derivative is consistent with unity, as expected from linear theory. At smaller scales and lower redshifts, however, it can pick up a sizeable enhancement which improves the constraints from cosmic shear. The vertical dashed lines indicate the approximate maximum wavenumber probed $k_{\rm max}\simeq \ell_{\rm max}/{\chi}$ assuming $\ell_{\rm max}=2000.$ The impact of baryons on the logarithmic derivative could be a significant systematic~\cite{Barreira:2019ckp}, which should be explored in future studies.

\end{document}